# Interpreting Purcell Enhancement of Non-Hermitian Metasurfaces with Spectral Parameters

Joshua T. Y. Tse*, Shunsuke Murai, and Katsuhisa Tanaka

Department of Material Chemistry, Graduate School of Engineering, Kyoto University, Katsura, Nishikyo-ku, Kyoto 6158510, Japan

**ABSTRACT**

The Purcell effect describes the enhancement of the spontaneous emission rate of an emitter near a resonant structure. However, evaluating the Purcell factor quantitatively and empirically is difficult due to the difficulties in measuring the electromagnetic nearfield of an optical resonance for calculation of the exact effective modal volume, especially with non-Hermitian resonators. Therefore, we propose a new analytical approach to circumvent the need to measure the nearfield and predict the Purcell enhancement with spectral parameters, which can be directly measured in farfield or fitted from such spectral measurements. Our proposed model predicts the averaged Purcell enhancement by metasurfaces on a photoluminescent medium, and is verified with experimental measurements and numerical simulations of nanoparticle arrays coupled to a fluorescent thin film. The model directly analyzes the photoluminescence enhancement and extraction efficiency of metasurface, and can be generalized to work with arbitrarily-shaped photoluminescent medium that is coupled to a resonator. This discovery provides a practical and accessible way to understand the underlying mechanisms of photoluminescence enhancement and will facilitate optimization of metasurfaces for efficient extraction of the enhanced luminescence.

*Email: tse@dipole7.kuic.kyoto-u.ac.jp

## I. INTRODUCTION

Purcell effect describes the enhancement of spontaneous emission of quantum emitters inside an optical cavity.[1-5] The spontaneous emission rate is increased proportionally by the increased local density of states (LDOS) within the optical cavity compared to free-space. Optical cavities can also function to redirect the emitted photons to specific out-coupling channels for various applications.[6-9] Metasurfaces and photonic crystals are commonly used to modify the LDOS near emitters in different coherent and incoherent light applications, such as fluorescent bio-imaging,[10-11] single-photon sources,[12-14] and nanolaser cavities.[15-16] To facilitate the design and optimization of efficient metasurfaces and photonic crystals, a comprehensive model that can provide physical insights for optimization is essential. A model that directly works with the empirical systems also alleviates the need for extensive computational time and thus would be preferable. Ideally, the model should also be compatible with both open cavities and dissipative materials, such that it can describe the out-coupling of the enhanced light as well as to be compatible with lossy plasmonic materials.

The maximal enhancement to spontaneous emission in an optical cavity is known as the Purcell factor $P_f$, which is given by:[1]

$$P_f = \frac{3}{4\pi^2}\left(\frac{\lambda}{n}\right)^3 \frac{Q}{V}$$

(1)

where $n$ is the refractive index in the cavity, $\lambda$ is resonant wavelength, $Q$ is the quality (Q) factor and $V$ is the effective modal volume of the resonance. Purcell's equation predicts that the Q-factor and the effective modal volume, indicating the temporal and spatial confinement of the cavity respectively, primarily contributes to the enhancement in the spontaneous emission rate. While the Q-factor is straightforward to measure as it is the ratio between the resonant frequency and the decay rate of the resonance, the effective modal volume is difficult to determine empirically or analytically for most optical cavities. For Hermitian systems, the effective modal volume at the resonant frequency is given by $V = [\int \varepsilon'(\mathbf{r})|\mathbf{E}(\mathbf{r})|^2 d^3r]/n^2$, where the nearfield of the cavity eigenmode $\mathbf{E}(\mathbf{r})$ is normalized to the maximum of $|\mathbf{E}(\mathbf{r})|$, and $\varepsilon'(\mathbf{r})$ is the relative permittivity in the resonator.[17-18] Similar formulations of the effective modal volume were also found for non-Hermitian systems. One approach is to modify the Hermitian formula with a surface integral term to capture the radiating field of the non-Hermitian system. This method was first developed by Leung et al.[19-20] and subsequently developed by Kristensen et al.[21-22] and Muljarov et al.[23] to provide an exact solution for the effective modal volume. Alternatively, Sauvan et al.[24-25] proposed a method which introduces artificial PMLs (perfectly matched layers) to convert the radiative loss into absorptive loss at the PMLs, which provides a finite integration volume to calculate the EM field of the resonant mode and subsequently the mode volume. While these computational approaches provided a method to bridge Purcell's model with non-Hermitian systems and improved our understanding of the nature of quasi-normal modes supported on non-Hermitian systems, a convenient method to evaluate the Purcell enhancement on experimental structures is still missing.

Therefore, we approach this problem by circumventing the need to acquire the exact effective modal volume and use spectral parameters that are related to the averaged Purcell enhancement directly instead. We make use of the formalism of the temporal coupled-mode theory (CMT), which describes optical resonators as 0-dimensional oscillators, to condense the nearfield spatial information into specific parameters that determine the spectral response of the resonator.[26-27] We choose this approach because while the Purcell factor defines the maximum enhancement, emitters are not always positioned at the antinode, and the averaged enhancement is a more practical quantity to be analyzed in most applications. In this work, we propose a unique method to understand the Purcell effect through such spectral parameters. The proposed model was derived based on the Lorentz reciprocity theorem and utilizes the parameters from a modified CMT derived from our previous work. We then proceed to verify our model with numerical simulation results based on the finite-difference time-domain (FDTD) method and discuss the application on experimental analysis. Lastly, we test the extension of the predictive power of our model by exploiting the similarities in different systems to reduce the spectral parameters needed.

## II. ANALYTICAL MODELLING OF PURCELL ENHANCEMENT

In our recent work, we explored the effectiveness of using metasurfaces made of different materials in enhancing fluorescent absorption by analyzing the competing absorption between plasmonic absorption and photoluminescence.[28] In particular, we discovered that the nearfield enhancement in a dye-embedded homogeneous layer can be mapped to the absorptive decay rate contributed by dye $\Gamma_{\text{abs,dye}}$, which is a parameter that is obtained through spectral measurement and fitting. As discussed in the previous work, $\Gamma_{\text{abs,dye}}$ is related to the nearfield enhancement by:

$$\frac{\Gamma_{\text{abs,dye}}}{\kappa} \propto \frac{\int_{\text{dye}} |\mathbf{E}(\mathbf{r})|^2 d^3r}{\int_{\text{all}} \varepsilon'(\vec{r}) |\mathbf{E}(\mathbf{r})|^2 d^3r}$$

(2)

where $\kappa$ is the extinction coefficient of the dye layer and $\varepsilon'(\mathbf{r})$ is the real part of the relative permittivity, the $\int_{\text{dye}} d^3r$ denotes a definite integral over the volume defined by the dye layer while $\int_{\text{all}} d^3r$ denotes an improper integral over all space. The fraction on the right-hand side compares the $|\mathbf{E}|^2$ inside the absorptive dye layer to the total $|\mathbf{E}|^2$ of the resonance mode at resonant frequency $\omega_0$. On the other hand, the Lorentz reciprocity theorem is commonly used in numerical analysis to predict the enhancement in fractional LDOS from the nearfield enhancement from a reciprocal source.[29-30] The Lorentz reciprocity theorem states that the EM field coupling is identical when we reverse the roles of source and antenna, and thus we can consider them as ports without the distinction between source and antenna. If we consider a quantum emitter with dipole moment **p** positioned at **r** and the planewave with momentum $\hbar \mathbf{k}$ as the two ports, we find that the probability for the dipole to emit a photon with energy $\hbar\omega$ that propagates away at $\hbar \mathbf{k}$ is proportional to the $|\mathbf{E}(\mathbf{r},\omega) \cdot \hat{\mathbf{p}}|^2$ when the resonance is excited by a plane wave source of $-\hbar \mathbf{k}$, or in other words the fractional LDOS is proportional to $|\mathbf{E}(\mathbf{r},\omega) \cdot \hat{\mathbf{p}}|^2$. The total emission intensity of an ensemble of randomly distributed and oriented dipole emitters thus can be evaluated through the integral $\int |\mathbf{E}(\mathbf{r},\omega)|^2 d^3r$ over the space occupied by the dipole emitters. Therefore, the photoluminescence enhancement (PLE) given by an optical resonance can be calculated by:

$$PLE(\omega) = \frac{\int_{\text{dye}} |\mathbf{E}(\mathbf{r},\omega)|^2 d^3r}{\int_{\text{dye}} |\mathbf{E}_{\text{ref}}(\mathbf{r},\omega)|^2 d^3r}$$

(3)

where the $|\mathbf{E}|^2$ of the resonance inside the dye layer is compared to the $|\mathbf{E}_{\text{ref}}|^2$ of the unpatterned structure to calculate the enhancement in fractional radiative LDOS, and thus the PLE. We combine these two insights to derive the following equation that predicts the PLE solely with spectral parameters (for details of the derivation see Appendix A):

$$PLE(\omega_0) = \frac{c\Gamma_{\text{rad}}}{\omega_0 t \Gamma_{\text{tot}}^2} \frac{\Gamma_{\text{abs,dye}}}{\kappa}$$

(4)

where $\omega_0$ is the resonant frequency, $\Gamma_{\text{rad}}$ is the radiative decay rate of the resonance, $t$ is the dye layer thickness, and $c$ is the speed of light in vacuum. $\Gamma_{\text{tot}}$ is the total decay rate and $\Gamma_{\text{tot}} = \Gamma_{\text{rad}} + \Gamma_{\text{abs,dye}} + \Gamma_{\text{abs,NP}}$, where $\Gamma_{\text{abs,NP}}$ is the absorptive decay rate contributed by the nanoparticles. (The decay rates are defined as the reciprocal of lifetime of the resonance.)

To better understand Eq. (4), we can group the terms based on where it originated during the derivation and we get $PLE(\omega_0) = (\Gamma_{\text{rad}}\Gamma_{\text{abs,dye}}/\Gamma_{\text{tot}}^2)(c/\omega_0 \kappa t)$, in which the first term $\Gamma_{\text{rad}}\Gamma_{\text{abs,dye}}/\Gamma_{\text{tot}}^2$ resembles the absorptivity by dye under resonance and the inverse of the second term $\omega_0 \kappa t/c$ resembles the absorption of an unpatterned dye layer at the limit of $\kappa \to 0$. This means that the Purcell enhancement is predicted to be the same as the absorption enhancement at $\omega_0$ and under the limit of $\kappa \to 0$. This is coherent with the fact that we used reciprocity to connect parameters from the absorption process to the analysis of the emission enhancement process. We can also compare Eq. (4) with Eq. (1) to understand the connection with Purcell's equation. By re-grouping

Eq. (4) into $PLE(\omega_0) = (\omega_0/\Gamma_{\text{tot}})[\Gamma_{\text{abs,dye}}/(\kappa t)](\Gamma_{\text{rad}}/\Gamma_{\text{tot}})(c/\omega_0^2)$, we can see that the first term $\omega_0/\Gamma_{\text{tot}}$ is the Q-factor. The second term $\Gamma_{\text{abs,dye}}/(\kappa t)$ is related to the effective modal volume $V$. From Eq. (2), $\Gamma_{\text{abs,dye}}/\kappa \propto \int_{\text{dye}} |\mathbf{E}(\mathbf{r})|^2/\left[\int_{\text{all}} \varepsilon'(\mathbf{r})|\mathbf{E}(\mathbf{r})|^2 d^3r\right] d^3r$, and the integrand $|\mathbf{E}(\mathbf{r})|^2/\left[\int_{\text{all}} \varepsilon'(\mathbf{r})|\mathbf{E}(\mathbf{r})|^2 d^3r\right] \propto 1/V$ when the numerator is evaluated at the maximum of $|\mathbf{E}(\mathbf{r})|$. Therefore, $\Gamma_{\text{abs,dye}}/(\kappa t)$ behaves similarly to an average of $1/V$ over the dye region as the PLE evaluates the effect over the whole dye layer instead of a single point in space. The third term $\Gamma_{\text{rad}}/\Gamma_{\text{tot}}$ defines the out-coupling efficiency of the optical cavity and the last term includes the remaining dimensional constants. This means that we can understand the $PLE(\omega_0)$ as the averaged Purcell enhancement factor over the specific spatial region covered by the definition of $\Gamma_{\text{abs,dye}}$.

It is also worth noting that this formulation is not limited to metasurfaces that is coupled to a flat dye layer and can be generalized to describe the Purcell enhancement on arbitrarily-shaped dye medium. Since $\Gamma_{\text{abs,dye}}/\kappa$ is characterized by the integral limits $\int_{\text{dye}} d^3r$, the shape of the volume of space defined by *dye* is not limited to a flat layer but can be generalized to any arbitrary shape. The thickness $t$, which is defined by $\int_{\text{dye}} 1/(\text{unit area})\, d^3r$, can also be generalized to the average thickness of the arbitrarily-shaped dye medium, or in other words the volume of *dye* per unit area on the surface. By treating the *dye* as an arbitrarily-shaped medium, this analysis can be generalized to other types of metasurfaces or photonic crystals. While the equation with $t$ may need some modifications to be expressed in terms of other variables to better suit the specific geometry in question, the principles and the derivation as outlined in Appendix A should remain as a good guideline in predicting the Purcell enhancement provided by the metasurface.

### III. NUMERICAL VERIFICATION WITH SURFACE LATTICE RESONANCE

We verified the proposed model by numerical simulations of metasurfaces that support surface lattice resonance (SLR), which is schematically illustrated in the inset of Fig. 1. Nanocylinders with height $H$ and diameter $D$ are placed in a square or hexagonal lattice with periodicity $P$. The nanocylinders are either composed of Al, Ag or TiO$_2$, where the refractive index of Al and Ag are obtained from Ref. [31] and the TiO$_2$ is set to have refractive index $n = 2.7$.[32] The geometrical parameters of the nanocylinders and the materials used are summarized in Appendix B. The lattice is placed on a silica glass substrate with refractive index $n = 1.46$ and embedded into an index-matching layer with thickness $t = 280$ nm and complex refractive index $n = 1.46 + i\kappa$. The extinction coefficient $\kappa$ is introduced to simulate the absorption of the dye layer at various dye concentrations. A semi-infinite layer of vacuum is placed above the index-matching layer. The $x$- and $y$-boundaries of the unit cell were terminated by periodic boundary conditions to simulate an infinite array, while PMLs were used to terminate the z-boundaries to simulate semi-infinite extension of the substrate and superstrate. A broadband plane wave source incident from the superstrate side was used to simulate the reciprocal of the photoluminescence towards the superstrate side. The transmissivity $T$ and reflectivity $R$ of the metasurface were recorded while the absorptivity $A$ was calculated by $A = 1 - T - R$. The $T$, $R$, $A$ spectra were fitted to obtain the parameters used in Eq. 4.[28,32] The local E-field within the index-matching layer was also recorded for calculating the numerical prediction of the PLE by Eq. 3 directly.

The comparison between the numerically simulated value of PLE and the PLE predicted by our proposed model is plotted in Fig. 1. The dashed line shows the diagonal where the predicted PLE and the simulated PLE matches. As illustrated in Fig. 1, the proposed model accurately predicts the FDTD simulated PLE. By comparing the data of different materials, we can see that the data for $TiO_2$ generally exhibit a larger PLE despite typically having weaker nearfield confinement than the plasmonic counterparts.[28] This can be explained by the much higher Q-factor of the $TiO_2$ metasurfaces. Since the $\Gamma_{tot}^2$ in the denominator is the only squared term in Eq. 4 and the Q-factor difference between the metasurfaces used is dominated by the decay rate, the $\Gamma_{tot}^2$ term exhibits the strongest effect in modulating the PLE strength of the metasurfaces.

## IV. EXPERIMENTAL VERIFICATION AND PARAMETER EXTENSION

We further test our model with experimentally measured PLE from the same type of metasurfaces that support SLR. $TiO_2$ nanocylinders with height $H$ = 90 nm and diameter $D$ = 130 nm are positioned in a square lattice with periodicity $P$ = 380, 400, 410, 420, and 430 nm (namely P380, P400, P410, P420 and P430) on a $SiO_2$ glass substrate (see Appendix C for details of sample preparation). The SEM image of P380 is shown in the inset of Fig. 2(c). The arrays are covered by a PMMA index-matching layer with thickness $t$ = 667 nm, which also contains a Lumogen dye (Lumogen F 305 red) of concentration $\rho$ = 1 or 2 wt% (weight percent).

The experimental setup for emission measurement is shown in Fig. 2(a). The Lumogen dye in the PMMA film was optically excited by a 445 nm blue laser from the substrate side and the photoluminescence intensity ($I$) was measured at the superstrate side and normalized against the emission intensity of the unstructured layer ($I_0$) to obtain the PLE ($I/I_0$). Fig. 2(b) – (c) shows the s-polarized PLE bandstructure and PLE spectrum of P380 ($\rho$ = 1 wt%) at 0°. In order to isolate the Purcell enhancement from the absorption enhancement effect, the PLE is renormalized such that the PLE baseline, indicated by the dashed line in Fig. 2(c), becomes unity. The physical meaning behind the renormalization is that we assume the Purcell enhancement is unity when detuned from all resonances and the absorption enhancement only affects the intensity of the emission spectrum but not the shape. The *s*-polarized PLE bandstructure ($I/I_0$) of other experimental samples are also measured and plotted in Fig 3.

Due to the use of reciprocity in the construction of our model, the working range in principle is limited to the overlapping wavelength range of the absorption and emission of the photoluminescence emitter, which only occurs from around 560 – 620 nm for the Lumogen dye in use (see Appendix D). The absorption range of the Lumogen dye limits the range where the factor $\Gamma_{\text{abs,dye}}/\kappa$ is well-defined and can be fitted, and the emission range of the Lumogen dye limits where the PLE is measurable. To overcome such limitations, we extend our model by borrowing parameters from other similar structures. In particular, the sample P380 with resonant wavelength (580 nm) within the overlapping range was used as a pivot and the factor $\Gamma_{\text{abs,dye}}/\kappa$ was borrowed to predict PLE of other samples with different $P$ but otherwise the same structure. The calibration of $\kappa$ of the index-matching layer is covered in Appendix E. The factor $\Gamma_{\text{rad}}/\Gamma_{\text{tot}}$ was also borrowed from P380 with $\rho = 0$ wt%, under the assumption that the effect of random scattering due to sample roughness is similar on all samples, since $\Gamma_{\text{tot}} = \Gamma_{\text{rad}} + \Gamma_{\text{scat}}$ when there is no absorption and $\Gamma_{\text{scat}}$ is the random scattering-contributed decay rate. The remaining terms are $c/\omega_0 t \Gamma_{\text{tot}}$ and the values of $\omega_0$ and $\Gamma_{\text{tot}}$ fitted from the PLE spectra of each sample respectively was used.

Fig. 4 shows the measured PLE compared to the PLE predicted by our proposed model using the aforementioned scheme, with the diagonal indicated by the dashed line. As shown in Fig. 4, the predicted PLE is consistent with the measured PLE. The general trend shows that the samples with larger $P$ exhibit higher PLE, which is consistent with the general increasing trend of the Q-factor as the SLR red-shifts and resembles closer to the Rayleigh anomaly.[33] On the other hand, we observe that the measured PLE of P400 is consistently lower than that of the prediction of our model. This can be explained as we compare the resonant wavelength of the SLRs supported on each sample to the absorption spectrum of the Lumogen dye. Due to the resonant wavelength of the SLR supported on P400 (610 nm) still being inside the absorption band of the Lumogen dye, the $\Gamma_{\text{tot}}$ of P400 is influenced by the dye absorption and our prediction scheme overestimated $\Gamma_{\text{rad}}/\Gamma_{\text{tot}}$ of P400, thus leading to a larger predicted value of PLE than the true measured value.

V. CONCLUSIONS

In conclusion, in order to circumvent the difficulty in calculating the exact effective modal volume of optical resonators in quantitative analysis of the Purcell effect, we proposed an analytical method that uses spectral parameters to model the Purcell effect instead. We derived the proposed model based on the Lorentz reciprocity theorem and utilized parameters from a modified CMT to represent the Purcell enhancement of optical resonances coupled to a dye layer. The derived equation expresses the averaged Purcell enhancement over the spatial region specified by the dye layer through connections with the absorption enhancement. We verified the proposed model with FDTD simulations and experimental PLE measurements of metasurfaces that support SLR. The proposed model also exhibits the potential to be further generalized. The model is compatible with arbitrarily-shaped dye-medium that is coupled to a resonator, and the equations can be updated with a suitable substitution on the geometrical descriptions of the dye-medium and resonator. On the application to experimental analysis, the limitation from the working range of fluorescent dye molecules was also overcame by extending certain parameters to similarly structured systems. Therefore, our new model provides direct analysis on the photoluminescence enhancement and extraction efficiency of experimental metasurfaces. It also exhibits sufficient flexibility to facilitates optimization of different metasurfaces efficiently.

## ACKNOWLEDGEMENTS

Financial support from Kakenhi (23K23044, 22K18884) is cordially acknowledged.

## APPENDIX A: DERIVATION OF THE PROPOSED MODEL

In our previous work, we discovered that $\frac{\Gamma_{\text{abs,dye}}}{\kappa} \propto \frac{\int_{\text{dye}} |\mathbf{E}(\mathbf{r})|^2 d^3r}{\int_{\text{all}} \varepsilon'(\mathbf{r}) |\mathbf{E}(\mathbf{r})|^2 d^3r}$.[28] Instead of a proportionality relation, we can convert this back to an equation by considering the power absorbed $P_{\text{abs}} = \Gamma_{\text{abs}} |a(\omega_0)|^2$, where $a$ is the mode amplitude used in the temporal coupled mode theory formalism and $|a(\omega_0)|^2$ is normalized to the total optical energy stored in the mode at resonant frequency $\omega_0$. This gives the equation:

$$\Gamma_{\text{abs}} = \frac{P_{\text{abs}}}{|a|^2} = \frac{\int \frac{1}{2}\omega_0 \varepsilon_0 \varepsilon''(\mathbf{r},\omega_0) |\mathbf{E}(\mathbf{r},\omega_0)|^2 d^3r}{|a(\omega_0)|^2}$$

(A1)

where the $P_{\text{abs}}$ is expressed in terms of the nearfield through the average dissipative energy density, $\varepsilon_0$ is the permittivity of vacuum and $\varepsilon''(\mathbf{r},\omega_0)$ is the imaginary part of the relative permittivity.[34-35] We can then confine the integration domain to the dye layer to derive:

$$\Gamma_{\text{abs,dye}} = \frac{\int_{\text{dye}} \frac{1}{2}\omega_0 \varepsilon_0 \varepsilon''(\mathbf{r},\omega_0) |\mathbf{E}(\mathbf{r},\omega_0)|^2 d^3r}{|a(\omega_0)|^2}$$

(A2)

Since the dye layer is a homogeneous medium, $\varepsilon''(\mathbf{r},\omega_0)$ becomes a constant over the integration space and can be factored out of the integral, giving:

$$\int_{\text{dye}} |\mathbf{E}(\mathbf{r},\omega_0)|^2 d^3r = \frac{2|a(\omega_0)|^2 \Gamma_{\text{abs,dye}}}{\omega_0 \varepsilon_0 \varepsilon''_{\text{dye}}(\omega_0)}$$

(A3)

This gives us the numerator in Eq. (3) and we can proceed to consider the denominator in Eq. (3). Since the reference structure is simply the dye layer on a transparent substrate, $\mathbf{E}_{\text{ref}}(\mathbf{r},\omega)$ can be derived directly from the incident by:

$$\langle s_+ | s_+ \rangle = \frac{1}{2}\sqrt{\frac{\varepsilon_0 \varepsilon}{\mu_0 \mu}} |\mathbf{E}_{\text{ref}}|^2 \cdot \text{(unit area)}$$

(A4)

where $|s_+\rangle$ is the incident wave vector with $\langle s_+ | s_+ \rangle$ normalized to the incident power. Eq. (3) thus can be expressed as:

$$PLE(\omega_0) = \frac{2|a(\omega_0)|^2 \Gamma_{\text{abs,dye}}}{2\langle s_+ | s_+ \rangle \sqrt{\frac{\mu_0 \mu_{\text{dye}}}{\varepsilon_0 \varepsilon_{\text{dye}}}} t \omega_0 \varepsilon_0 \varepsilon''_{\text{dye}}(\omega_0)}$$

(A5)

where $t = \int_{\text{dye}} 1/(\text{unit area})\, d^3r$ is the thickness of the dye layer. The relationship between the $|a(\omega)|^2$ and the $\langle s_+ | s_+ \rangle$ can be derived through the CMT formalism, in which the steady-state solution is given by: $|a(\omega)|^2 = \frac{\frac{\Gamma_{\text{rad}}}{2} |\langle v^* | s_+ \rangle|^2}{(\omega - \omega_0)^2 + \left(\frac{\Gamma_{\text{tot}}}{2}\right)^2}$, where $|v^*\rangle$ is the in-coupling constant with the normalization $\langle v | v \rangle = 2$.[27-28] After simplifying the constants in Eq. (A5), the PLE is derived to be:

$$PLE(\omega_0) = \frac{c \Gamma_{\text{rad}}}{\omega_0 t \Gamma_{\text{tot}}^2} \frac{\Gamma_{\text{abs,dye}}}{\kappa} \frac{|\langle v^* | s_+ \rangle|^2}{\langle s_+ | s_+ \rangle}$$

(A6)

Since we are evaluating the PLE at a single port under normal emission, $|\langle v^* | s_+ \rangle|^2 = \langle s_+ | s_+ \rangle$ and we can simplify the PLE of the metasurface to be:

$$PLE(\omega_0) = \frac{c \Gamma_{\text{rad}}}{\omega_0 t \Gamma_{\text{tot}}^2} \frac{\Gamma_{\text{abs,dye}}}{\kappa}$$

(A7)

**APPENDIX B: GEOMETRICAL PARAMETERS FOR FDTD SIMULATION**

The geometrical parameters of the nanocylinders and the materials used in FDTD simulations are summarized in Table B1.

**APPENDIX C: SAMPLE PREPARATION**

The TiO$_2$ metasurfaces are prepared on a silica glass substrate by electron-beam lithography and reactive ion etching.[32] First, a TiO$_2$ thin layer with thickness of 90 nm was deposited on the SiO$_2$ glass substrate by RF (radio frequency) magnetron sputtering. The X-ray diffraction (XRD) pattern of the TiO$_2$ layer is measured after sputtering. As shown in Fig. C1, the absence of distinct peaks indicates that the TiO$_2$ layer is amorphous. Then, a resist (ZEP520A) was spin-coated and nanohole array patterns were written by electron-beam lithography. After that, a Cr layer (120 nm) was deposited by electron-beam deposition, and the following lift-off process resulted in Cr dot array patterns on the TiO$_2$ layer. Then, the TiO$_2$ layer was etched by reactive ion etching with CHF$_3$ gas to make the nanocylinder arrays. The Cr dot mask was then removed by wet etching. A PMMA index-matching layer that contains Lumogen F 305 red dye is then spin coated onto the sample.

**APPENDIX D: ABSORPTION AND EMISSION SPECTRUM OF LUMOGEN DYE**

The absorption and emission spectra of the Lumogen F 305 red dye were measured as a reference. A PMMA layer with embedded Lumogen dye was spin coated on an unstructured SiO$_2$ glass substrate. Fig. D1 shows the measured absorption and emission of the Lumogen dye normalized to the respective absorption and emission peak values. As highlighted in Fig. D1, the overlapping region of the absorption and emission of the Lumogen dye only exists around 560 – 620 nm.

**APPENDIX E: CALIBRATION OF $\kappa$ IN THE EXPERIMENTAL SAMPLE**

We measured the extinction coefficient $\kappa$ of the PMMA layer with the Lumogen dye. The normalized transmissivity of the dye layer with $\rho$ = 1 wt% was measured by UV/VIS/NIR spectroscopy and was fitted by the Fresnel's Equations. The permittivity of the dye layer was modelled by the generalized Lorentz oscillator model with 5 oscillators.[36] The transmissivity is plotted in Fig. E1(a) with the best fit represented by the red line, and the fitted $n$ and $\kappa$ of the dye layer are plotted in Fig. E1(b). In particular, the $\kappa$ at the resonant wavelength of P380 was found to be $\kappa$ = 0.00533 while the thickness of the PMMA layer was fitted to be 667 nm. The $\kappa$ is then assumed to be proportional to $\rho$.

Figure 1. The PLE predicted by the proposed model is plotted against the FDTD simulated PLE. The dashed line shows the diagonal where the predicted PLE and the simulated PLE matches. Inset: The design of the metasurface that supports SLR. Nanocylinders with diameter $D$ and height $H$ are aligned in a 2D lattice of periodicity $P$ on a silica glass substrate (blue) and embedded into an index-matching layer (pink) with thickness $t$.

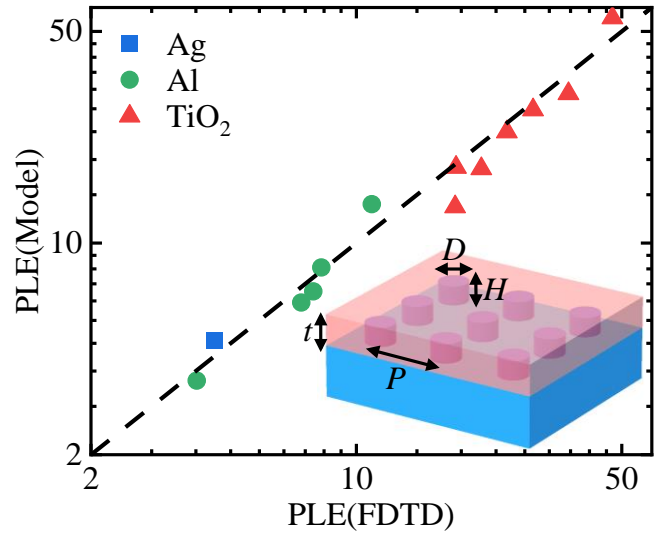

Figure 2. (a) The experimental setup for emission measurement is illustrated. A 445 nm laser coupled through an optical fiber was collimated by a lens (L1) and the excitation beam was controlled to be *s*-polarized by a linear polarizer (LP1). The incident angle was fixed at $\theta = 60°$. LP2 was used to select the emission polarization and L2 was used to focus the luminescence in the direction normal to the sample (S) into an optical fiber connected to a spectrometer. The detection arm was rotated around S to measure the emission at different angles. (b) The *s*-polarized PLE bandstructure ($I/I_0$) of sample P380. (c) The PLE spectrum ($I/I_0$) of P380 at emission normal to the surface of the nanoparticle array. The red dashed line indicates the PLE baseline for renormalization. The SEM image of sample P380 is shown in the inset. The scale bar is 500 nm.

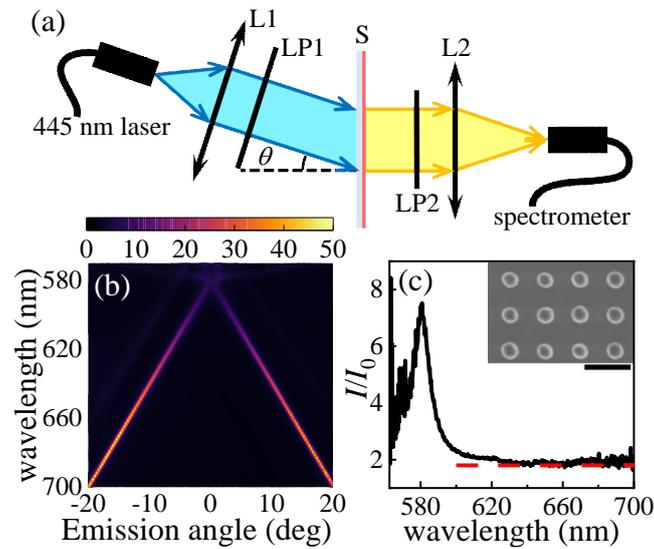

Figure 3. The *s*-polarized PLE bandstructure ($I/I_0$) of the samples (a) P400, (b) P410, (c) P420 and (d) P430.

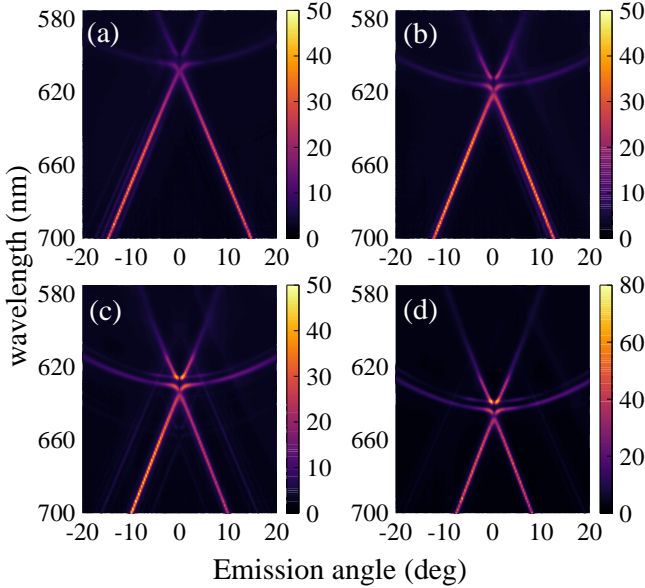

Figure 4. The PLE predicted by the model is plotted against the measured normalized PLE with $\rho =$ 1 and 2 wt% for each nanoparticle array. The dashed line shows the diagonal where the predicted PLE and the measured PLE matches each other.

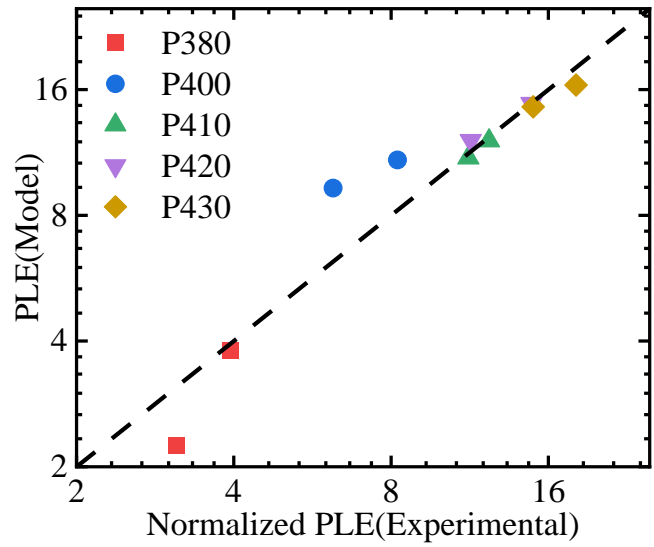

Table B1. The geometrical parameters of the structures used in FDTD simulations.

| $D$ (nm) | $H$ (nm) | $P$ (nm) | Lattice type | Material |
|---|---|---|---|---|
| 90 | 100 | 380 | square | Al |
| 80 | 100 | 380 | square | Al |
| 130 | 100 | 380 | square | Al |
| 90 | 50 | 380 | square | Al |
| 90 | 100 | 440 | hexagonal | Al |
| 90 | 100 | 380 | square | Ag |
| 130 | 100 | 380 | square | $TiO_2$ |
| 110 | 100 | 380 | square | $TiO_2$ |
| 150 | 100 | 380 | square | $TiO_2$ |
| 110 | 100 | 380 | hexagonal | $TiO_2$ |
| 130 | 100 | 380 | hexagonal | $TiO_2$ |
| 130 | 100 | 350 | square | $TiO_2$ |
| 130 | 100 | 300 | square | $TiO_2$ |

Figure C1. The XRD pattern of the as-deposited TiO$_2$ layer.

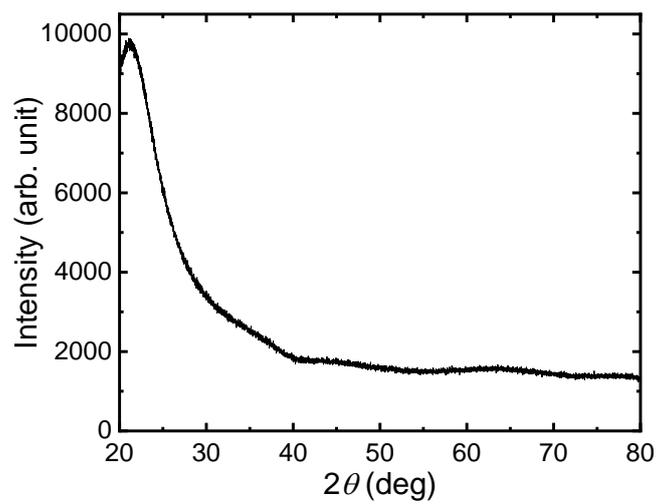

Figure D1. The normalized absorption and emission spectra of the Lumogen dye. The overlapping region from 560 – 620 nm is highlighted.

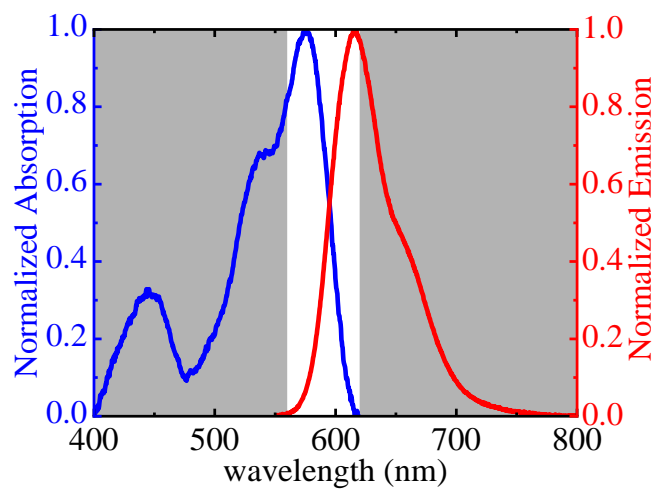

Figure E1. (a) The transmissivity of the Lumogen in PMMA dye layer with $\rho = 1$ wt%. The best fit is plotted as the red line. (b) The fitted $n$ and $\kappa$ of the dye layer.

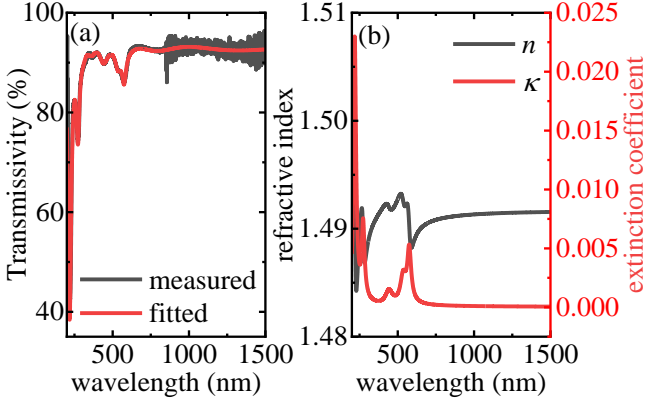